\documentclass{aa}  

\usepackage{txfonts,epsfig,graphicx,url,twoopt,amssymb}
\usepackage[breaklinks=true]{hyperref} 
\usepackage[usenames, dvipsnames]{color}
\hypersetup{colorlinks=true,citecolor=blue,linkcolor=blue,urlcolor=red}
\graphicspath{{./figures/}}
\begin{document}

   \title{How dust fragmentation may be beneficial to planetary growth by~pebble accretion}

   \author{J. Dr\k{a}\.{z}kowska$^1$
          \and
          S. M. Stammler$^1$
          \and
          T. Birnstiel$^{1,2}$
          }

   \institute{University Observatory, Faculty of Physics, Ludwig-Maximilians-Universit\"at M\"unchen, Scheinerstr.~1, 81679 Munich, Germany
             \and
             Exzellenzcluster ORIGINS, Boltzmannstr.~2, D-85748 Garching, Germany
             }

   \date{Received ; accepted}

 
  \abstract
   {Pebble accretion is an emerging paradigm for the fast growth of planetary cores. Pebble flux and pebble sizes are the key parameters used in the pebble accretion models.}
   {We aim to derive the pebble sizes and fluxes from state-of-the-art dust coagulation models, understand their dependence on disk parameters and the fragmentation threshold velocity, and the impact of those on the planetary growth by pebble accretion.}
   {We use a one dimensional dust evolution model including dust growth and fragmentation to calculate realistic pebble sizes and mass flux. We use this information to integrate the growth of planetary embryos placed at various locations in the protoplanetary disk.}
   {Pebble flux strongly depends on disk properties, such as its size and turbulence level, as well as on the dust aggregates fragmentation threshold. We find that dust fragmentation may be beneficial to planetary growth in multiple ways. First of all, it prevents the solids from growing to very large sizes, for which the efficiency of pebble accretion drops. What is more, small pebbles are depleted at a lower rate, providing a long lasting pebble flux. As the full coagulation models are computationally expensive, we provide a simple method of estimating pebble sizes and flux in any protoplanetary disk model without substructure and with any fragmentation threshold velocity.}
   {}


   \maketitle
%

\section{Introduction}

In the classical paradigm of planet formation a significant fraction of solids is very quickly converted into kilometer-sized planetesimals. Some of those planetesimals continue to grow rapidly via the runaway growth \citep[e.g.,][]{1989Icar...77..330W, 1993Icar..106..210I, 1996Icar..123..180K}. However, this fast stage of growth is soon over as the embryo starts to stir its feeding zone and the accretion transitions to oligarchic growth stage, slowing down with embryo mass \citep[e.g.,][]{1998Icar..131..171K, 2010ApJ...714L.103O}. By the end of the oligarchic growth, the embryo absorbs all of the material in its immediate vicinity and reaches the isolation mass, which is as low as Mars mass in the terrestrial planets region, but increases with the distance to the star \citep{2002ApJ...581..666K}. The isolation mass may be high enough to reproduce the cores of giant planets but the core growth timescale in the planetesimal-driven scenario becomes prohibitively long to allow for accretion of gaseous atmosphere outside of Jupiter location \citep{2002AJ....123.2862T, 2003Icar..161..431T, 2010AJ....139.1297L, 2019A&A...631A..70J}.

These drawbacks of the classical planetesimal-driven model motivated development of the alternative scenario called pebble accretion. In this paradigm, the embryo is accreting centimeter-sized pebbles rather than planetesimals. Gravity and gas drag act together to enhance the cross section of a planetary embryo for its encounters with pebbles and thus speed up the growth \citep{2010A&A...520A..43O, 2012A&A...544A..32L}. Because pebbles are rapidly drifting through the protoplanetary disk, the size of the feeding zone increases, allowing the embryos to grow to larger sizes before the pebble flux is halted by planet-disk interactions \citep{2014A&A...572A..35L}. 

In the pebble accretion scenario, the outcome of planet formation is determined by the sizes and flux of the pebbles drifting through the protoplanetary disk. \citet{2014AJ....148..109K} used large pebbles with a size distribution between 0.5~m and 5~m (corresponding to the Stokes number of 0.25 and 25) and a constant pebble flux of 0.13 M$_\oplus$~yr$^{-1}$. They found that reproducing the giant planets in the Solar system is challenging as pebble accretion tends to convert too many planetesimals into large embryos. \citet{2015Natur.524..322L} showed that this difficulty may be mitigated if pebbles form gradually so that the growing planetesimals have time to interact and stir each other. Recently, \citet{2019A&A...627A..83L} showed that a high pebble flux leads to formation of giant planets, while a lower flux leads to formation of super-Earths. However, many authors studying the pebble-driven planet growth for simplicity define a fixed size or Stokes number of pebbles and the value of the pebble flux (or the pebble to gas flux ratio) as arbitrary parameters (typically fixed in orbital distance and/or time), independent of the protoplanetary disk model \citep{2015SciA....1E0109J, 2019A&A...624A.114L, 2019A&A...627A..83L, 2019A&A...622A.202J, 2020ApJ...892..124O, 2020NatAs...4..492B, 2020MNRAS.tmp.1862W}.

\citet{LJ2014} presented an algorithm for calculating the pebble flux and pebble Stokes number, which is often used by the community \citep{2015Natur.524..322L, 2015PNAS..11214180L, 2016A&A...591A..72I, 2016A&A...590A.101B, 2017A&A...607A..67M, 2018A&A...619A.174B, 2019A&A...630A..51B, 2019A&A...623A..88B, 2019arXiv190208772I}. In this model, which is based on a simplified scenario of dust evolution, it is assumed that all dust grains grow until the growth timescale becomes longer than the drift timescale. Since the growth timescale strongly increases with distance, the "pebble formation front" moves outward, leading to subsequent portions of solids decoupling from the gas and maintaining inward pebble flux until the front reaches the outer edge of the disk. This picture is roughly consistent with the outcome of dust coagulation models as long as dust growth can proceed without being hindered by fragmentation. However, the increase of collision speeds with aggregate sizes often prevents dust growth to reach those sizes.

Laboratory experiments show that collisions of dust aggregates become destructive already at speeds much lower than predicted for the protoplanetary disk environment. The exact fragmentation threshold speed remains uncertain and it was shown to strongly depend on the porosity and composition of the aggregates \citep{2011ApJ...737...36W, 2013MNRAS.435.2371M, 2018SSRv..214...52B}. Until recently, it was believed that the icy aggregates outside of the water ice line are significantly more sticky than the dry aggregates inside of the ice line \citep{2014MNRAS.437..690A, 2015ApJ...798...34G}. However, newer laboratory data do not support this view, demonstrating that the icy aggregates break easily \citep{2018MNRAS.479.1273G, 2019ApJ...874...60S, 2019ApJ...873...58M}. 

Nevertheless, most of the works concerned with planet growth via pebble accretion published to date neglect the effect that fragmentation of pebbles has on their sizes and flux (the exceptions are \citealt{2016ApJ...825...63C, 2018ApJ...865...30C}, \citealt{2020A&A...642A.140G}, and \citealt{2020A&A...643L...1V, 2020arXiv200805497V}). In this paper, for the first time, we study the growth of planetary embryo by pebble accretion in connection with a self-consistent dust evolution model considering full size distribution obtained in detailed dust coagulation simulation. 

This paper is organized as follows. We present the numerical model used in this work in Sect.~\ref{sub:Model}. We present the resulting planet growth tracks and their dependence on the parameters that are the most important to defining dust evolution outcome in Sect.~\ref{sub:Results}. We present a simple way of predicting a realistic pebble flux and sizes which is valid in a fragmentation dominated as well as in drift dominated protoplanetary disk in Sect.~\ref{sub:predictor}. In Sect.~\ref{sub:Discussion} we discuss our work and we summarize it in Sect.~\ref{sub:Summary}.

\section{Model}\label{sub:Model}

\subsection{Protoplanetary disk model}

\begin{figure}
\centering
\includegraphics[width=\hsize]{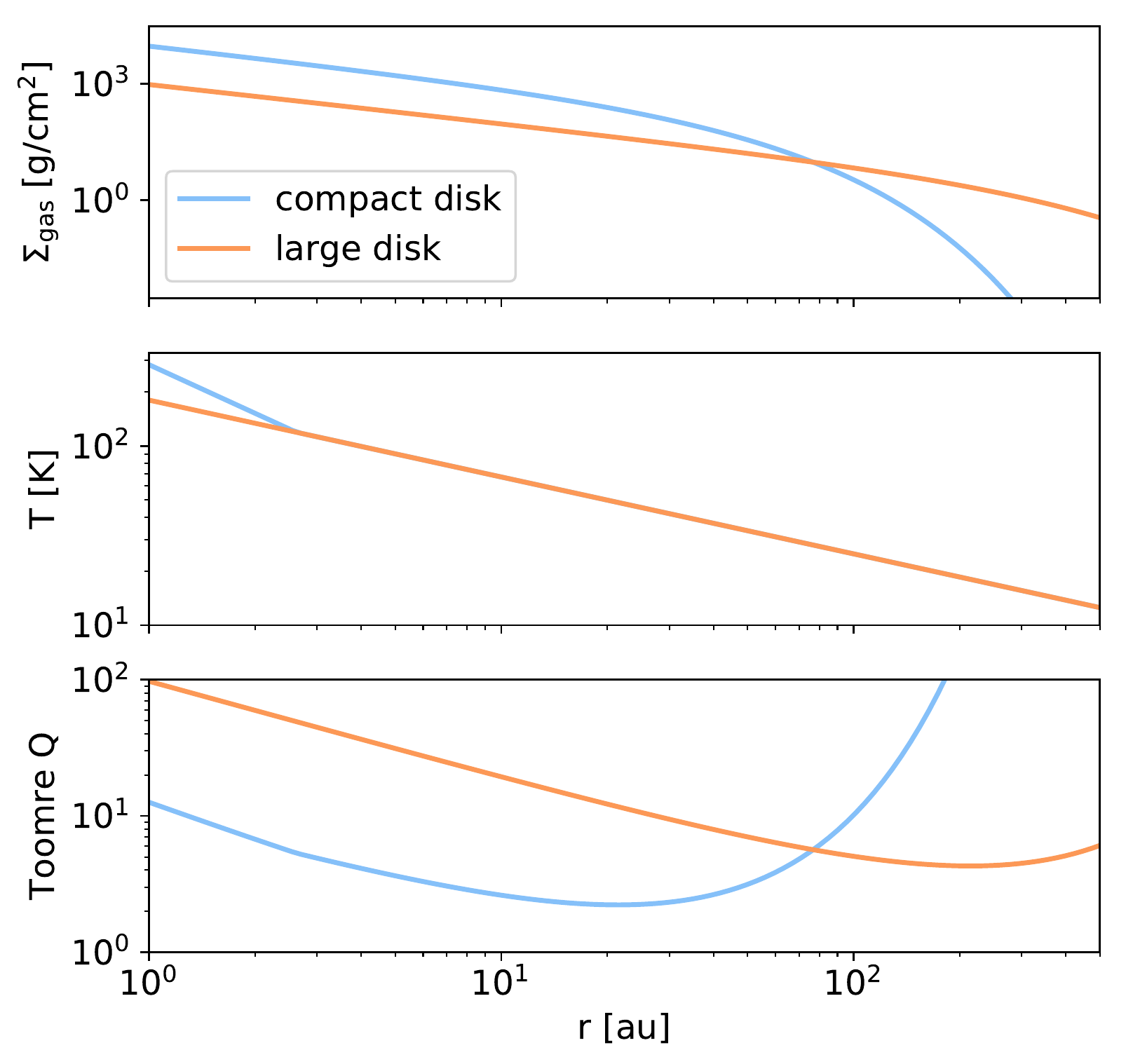}
\caption{Comparison of the compact and large disk models with $\alpha=10^{-4}$. {\it Top:} Gas surface density profiles. {\it Middle:} Temperature in disk midplane. {\it Bottom:} Toomre $Q$ parameter for both disk models used in this paper. }
         \label{fig:disks}
\end{figure}

\begin{figure*}
\centering
\includegraphics[width=\hsize]{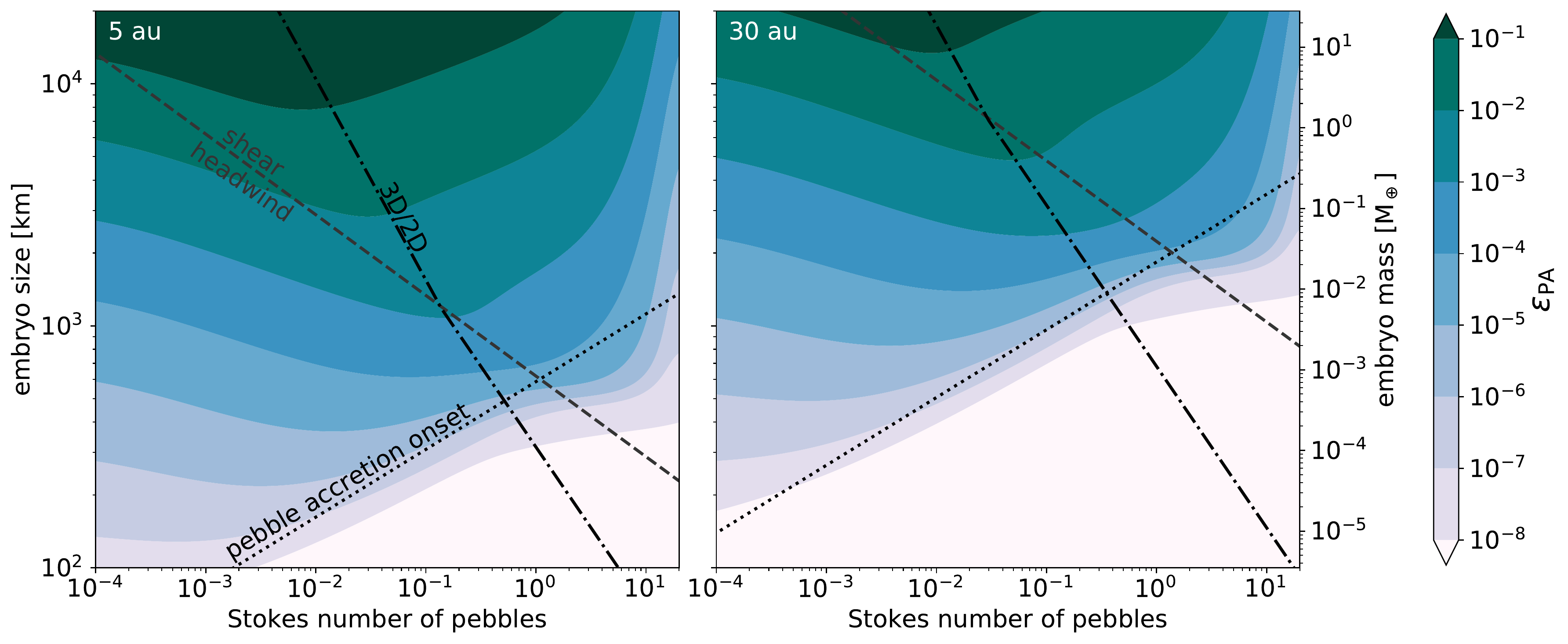}
\caption{The efficiency of pebble accretion parameter $\varepsilon_{\rm{PA}}$ as a function of the planetary embryo size and pebbles Stokes number calculated at 5~au ({\it left panel}) and at 30~au ({\it right panel}) in the compact disk model with $\alpha=10^{-4}$. The dotted line corresponds to the pebble accretion onset (when the aerodynamic effects become important). The grey dashed line shows the transition between the headwind and shear regimes of pebble accretion. The dashed-dotted line shows the transition between 3D and 2D pebble accretion.  }
         \label{fig:epsPA}
\end{figure*}

We model protoplanetary disk surrounding a solar mass star with effective temperature of  $T_{\rm{eff}}=5772$~K. In the initial condition, we set the gas surface density as a function of the radial distance $r$ to the often used, self-similar solution to the viscous evolution equations \citep{1974MNRAS.168..603L} 
\begin{equation}
    \Sigma_{\rm{gas},0} = \frac{M_{\rm{disk}}}{2 \pi r_c^2} \left(\frac{r}{r_c}\right)^{-1}\exp{\left(-\frac{r}{r_c}\right)},
\end{equation}
where $M_{\rm{disk}}$ is the initial disk mass, and $r_c$ is the characteristic radius, and the dust surface density to 
\begin{equation}
    \Sigma_{\rm{dust},0} = Z\cdot\Sigma_{\rm{gas},0},
\end{equation}
where $Z$ is the global solids-to-gas ratio. In all the models presented in this paper, we consider disks with exactly the same initial mass budget. $M_{\rm{disk}}$ is set to 0.2~M$_\odot$ and $Z$ has the standard value of 0.01, which corresponds to the initial mass in solids of about 650~M$_\oplus$. We consider two values of the characteristic radius, $r_c=30$~au in the compact disk model and $r_c=300$~au in the large disk model. The initial gas surface density profiles are presented in the upper panel of Fig.~\ref{fig:disks}. Despite their high mass, both of the disk models are gravitationally stable, with the Toomre $Q>2$, as shown in the bottom panel of Fig.~\ref{fig:disks}.

The gas disk evolves viscously, which means that the characteristic radius increases, while the surface density drops with time. The speed of viscous evolution is determined by the $\alpha$ parameter. We set the default value to $\alpha=10^{-4}$. In Section~\ref{sub:alpha}, we show the effects of varying the $\alpha$ parameter. In this work, we do not consider disk dispersal by photoevaporation. 

The pebble accretion rate is sensitive to the vertical scale height of gas $H_{\rm{gas}}$ and thus to the disk temperature structure. In this paper, we take into account both stellar irradiation and viscous heating using the simple prescription proposed by \citet[][their equation 7 and 8]{2016A&A...591A..72I}. Most of the disk is heated by stellar irradiation, leading to a shallow temperature profile $T_{\rm{irr}}\propto r^{-3/7}$. In the inner part of the disk, viscous heating may change the temperature profile to a steeper function of the radial distance, $T_{\rm{vis}}\propto r^{-9/10}$. In the viscous heating regime, the temperature depends on the gas mass accretion rate, and thus this effect is particularly important in the compact disk model, as visible in the middle panel of Fig.~\ref{fig:disks}, and in models with $\alpha=10^{-3}$, where the gas mass flux, defined as $\dot{M}_{\rm{gas}}=3\pi\alpha\Sigma_{\rm{gas}}H_{\rm{gas}}^2\Omega_{\rm{K}}$, with $\Omega_{\rm{K}}$ being the Keplerian frequency, is higher. 

\subsection{Dust evolution}

We investigate the dust evolution on a logarithmic radial grid with 120 grid points spaced between 1~au and 1000~au. We compute dust evolution in the azimuthally and vertically averaged framework using the \texttt{DustPy} code, a Python-based version dust coagulation model similar to the code presented by \citet{2010A&A...513A..79B}. At every radial distance to the star, dust mass grid with 9 mass bins per mass decade is built and the Smoluchowski equation is used to solve for dust growth and fragmentation. We assume that the dust grains initially have radius of one micron and internal density of 1.25~g~cm$^{-3}$. The collisions are driven by the Brownian motion, turbulence prescribed with the $\alpha$ parameter, radial and azimuthal drift, and dust settling. We consider various values of the fragmentation threshold speed $v_{\rm{f}}$, between 1~m/s and $\infty$, corresponding to no fragmentation at all. In line with the newest laboratory results, we do not employ a transition at the water ice line, but rather we use a single value of $v_{\rm{f}}$ independent on the radial distance.

Radial drift of particles in every mass bin $i$ is computed based on their local Stokes number \citep{1977MNRAS.180...57W} $\it{St}_i$:
\begin{equation}\label{eq:vr}
    v_{r,i} = \frac{2 v_\eta \it{St}_i}{1 + \it{St}^2_i} + \frac{v_{\rm{gas}}}{1 + \it{St}^2_i}, 
\end{equation}
where $v_{\rm{gas}}$ is the local gas velocity (driven by gas viscous evolution) and $v_\eta$ is the maximum velocity of the radial drift driven by the local pressure gradient $\partial_r P_{\rm{gas}}$. We calculate the $v_\eta$ as
\begin{equation}\label{eq:veta}
    v_\eta = \frac{\partial_r P_{\rm{gas}}}{2\rho_{\rm{gas}} \Omega_{\rm{K}}},
\end{equation}
where $\rho_{\rm{gas}}=\Sigma_{\rm{gas}}/(\sqrt{2\pi}H_{\rm{gas}})$ is the midplane density of gas.

We take into account both the Epstein and the Stokes drag regimes when calculating the Stokes number $\it{St}_i$. In this work, we do not include the backreaction from dust to gas.

Further details about the dust evolution code can be found in \citet{2010A&A...513A..79B} and in forthcoming Stammler \& Birnstiel (in prep.).

\begin{figure*}
\centering
\includegraphics[width=\hsize]{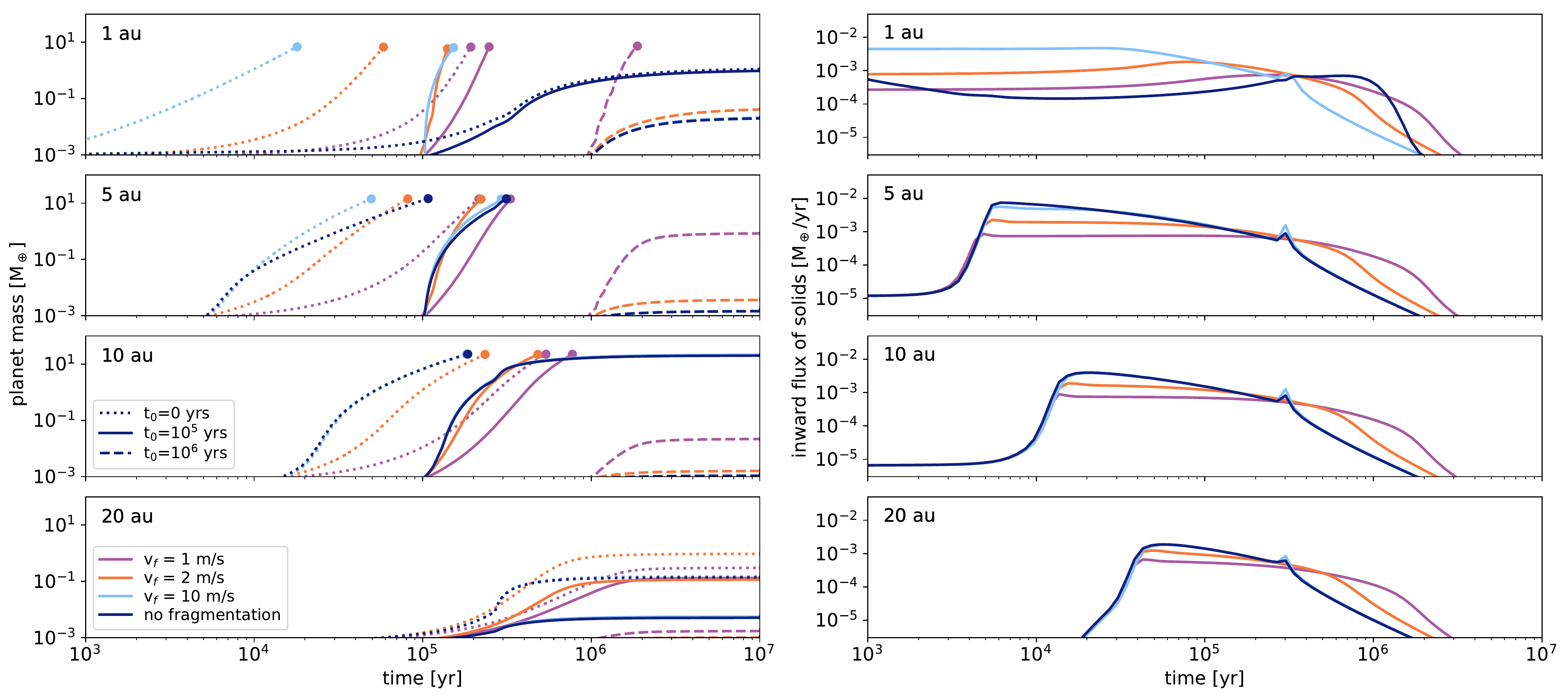}
\caption{{\it Left:} Planetary embryos growth via pebble accretion in the compact protoplanetary disk in models with $\alpha=10^{-4}$ with different fragmentation threshold speed and one not including dust fragmentation at all. In all cases where the line representing the v$_{\rm{f}}=10$~m/s run is not visible, it is identical with the "no fragmentation" run. Lines ended with a circle represent growth tracks that reached the pebble isolation mass. {\it Right:} Inward flux of solids (integrated all dust sizes) at different locations in the same models.}
         \label{fig:planets_fragmentation}
\end{figure*}

\subsection{Planet growth by pebble accretion}

We consider growth of planetary embryos with starting mass of 0.001~M$_\oplus$ (approximately 0.1 Moon mass, or 7 masses of the largest asteroid, Ceres). We model the growth of embryos at various radial locations and do not consider their migration or planet-disk interaction. We do not subtract the dust accreted by an embryo from the dust evolution code, which means that we assume that each model corresponds to a single planet. We consider various possibilities for the point at which we insert the embryos into the simulation $t_0$: either we let the pebble accretion start to grow from the beginning of the simulation, or after 10$^5$~years and after 10$^6$~years. We assume that the embryos are on circular orbits with zero inclination. 

To calculate the pebble accretion efficiency, we use the method provided by \citet{2018A&A...615A.138L} and \citet{2018A&A...615A.178O} who used 3-body calculations to obtain a general recipe for the pebble accretion efficiency of a single planet\footnote{Python implementation of this algorithm is publicly available at \href{url}{https://staff.fnwi.uva.nl/c.w.ormel/software/epsilon.tar.gz}}. In Fig.~\ref{fig:epsPA}, we show the example of pebble accretion efficiencies obtained using this recipe for a range of planetary embryo masses and sizes at 5~au and at 30~au in the initial condition of the compact disk model with the turbulence parameter $\alpha=10^{-4}$. As can be seen in this figure, the recipe takes into account both the 2D regime, where the embryo is large enough to accrete from the complete layer of pebbles, and the 3D regime, where the small embryo only has access to some fraction of the pebble layer. The transition from 3D to 2D regime is marked with the dashed-dotted line. The grey dashed line shows the transition between the headwind and shear regimes. The headwind regime, also called Bondi regime, is valid for small embryos or small pebbles, when the pebbles are accreted only from the embryo proximity and the relative velocity between the pebbles and the embryo is determined by the gas flow. Larger embryos enter the shear regime, also called Hill regime, when the pebbles are accreted from much larger area and their approach velocity is determined by the Keplerian shear. The transition between the headwind and shear regimes does not impact the pebble accretion efficiency in the 3D regime, while in the 2D regime it slightly changes the dependence of $\varepsilon_{\rm{PA}}$ on the Stokes number. This effect becomes more important in the outer regions of the disk, when the transition between headwind and shear regime shifts to higher embryo masses. As explained in \citet{2017ASSL..445..197O}, in the 3D regime the pebble accretion efficiency increases with the Stokes number because dust settling is more efficient and the embryo accretes from a denser midplane layer, while in the 2D regime this effect is canceled by the increasing drift speed of pebbles and the decreasing embryo-pebble interaction time. The dotted line in Fig.~\ref{fig:epsPA} shows the pebble accretion onset, that is the minimum embryo mass for which the aerodynamic deflection becomes important \citep{2016A&A...586A..66V}. For reference, the typical planetesimal size formed in the streaming instability models is about 100~km \citep{2017A&A...597A..69S, 2019ApJ...885...69L, 2020ApJ...901...54K}. Interestingly, these smallest embryos could only grow by accretion of small pebbles and with an extremely low efficiency. The pebble accretion onset shifts toward even larger embryo masses in the outer parts of the disk.

Figure~\ref{fig:epsPA} shows that the pebble accretion efficiency is mostly dependent on the embryo mass and not on the pebble size, as long as the Stokes number stays between $10^{-3}$ and 1, in the typical "pebble-size" regime. {A relatively large embryo, with significant fraction of an Earth mass, is able to grow quickly by accreting a considerable fraction of the pebble flux. It is worth noting that multiple small embryos could grow at the same time, collectively capturing similar fraction of pebbles as a single large embryo \citep{2019A&A...624A.109B}}. The growth rate of the planetary embryo depends both on the pebble accretion efficiency and the pebble flux. In this paper, we compute pebble fluxes consistently with the pebble size resulting from our full dust coagulation model.

Our pebble accretion algorithm works as follows. At the planetary embryo location $a_e$, we calculate the Stokes number of the dust particles from every mass bin, $\it{St}_i$. Then we calculate the pebble accretion efficiency $\varepsilon_{\rm{PA}}(m_e,a_e,\it{St}_i)$ for the current embryo mass $m_e$ and multiply it by the flux of solids corresponding to each bin of the mass grid $\dot{m}_{p}(a_e,\it{St}_i)$ obtained in the dust evolution code. In the next step, we sum the contributions from each dust mass bin to obtain the embryo growth rate
\begin{equation}\label{eq:planetgrowth}
    \dot{m}_{e} = \sum_i{\varepsilon_{\rm{PA}}\left(m_e,a_e,\it{St}_i\right)} \cdot \dot{m}_{p}(a_e,\it{St}_i).
\end{equation}

The growth stops when the embryo reaches the pebble isolation mass. Following \citet{2014A&A...572A..35L}, we calculate the pebble isolation mass as
\begin{equation}
    M_{\rm{iso}} = 20\ M_\oplus \cdot\left(\frac{H_{\rm{gas}}/r}{0.05}\right)^3,
\end{equation}
where $H_{\rm{gas}}/r$ is disk aspect ratio. It is worth noting that \citet{2018A&A...615A.110A} and \citet{2018A&A...612A..30B} reported that the pebble isolation mass may be increased in turbulent disks while \cite{2020A&A...638A..97Z} found that the pebble isolation mass is decreased when a realistic equation of state and radiative cooling are taken into account.

\section{Results}\label{sub:Results}

We performed seven different models, varying the fragmentation threshold velocity, disk size, and turbulence level. In each of the models, we considered growth of planetary embryos by pebble accretion at 1~au, 5~au, 10~au, 20~au, 30~au, 40~au, and 50~au. However, in none of the models did an embryo located outside of 20~au reach the pebble isolation mass.

\subsection{Fragmentation threshold}\label{sub:frag}

We first focus on the impact of the fragmentation threshold value on the planetary growth. Figure~\ref{fig:planets_fragmentation} presents results of four runs, all performed with the compact disk model with the turbulence strength $\alpha=10^{-4}$ and different fragmentation threshold speeds $v_{\rm{f}}=1$~m/s, $v_{\rm{f}}=2$~m/s, $v_{\rm{f}}=10$~m/s, and $v_{\rm{f}}=\infty$, corresponding to no fragmentation. The left panel of Fig.~\ref{fig:planets_fragmentation} presents the growth of planetary embryos at 1~au, 5~au, 10~au, and 20~au. At each location, we tested different embryo introduction times $t_0$: at the beginning of the simulation ($t_0=0$), after $10^5$~years, and after $10^6$~years of dust evolution. 

\begin{figure}
\centering
\includegraphics[width=\hsize]{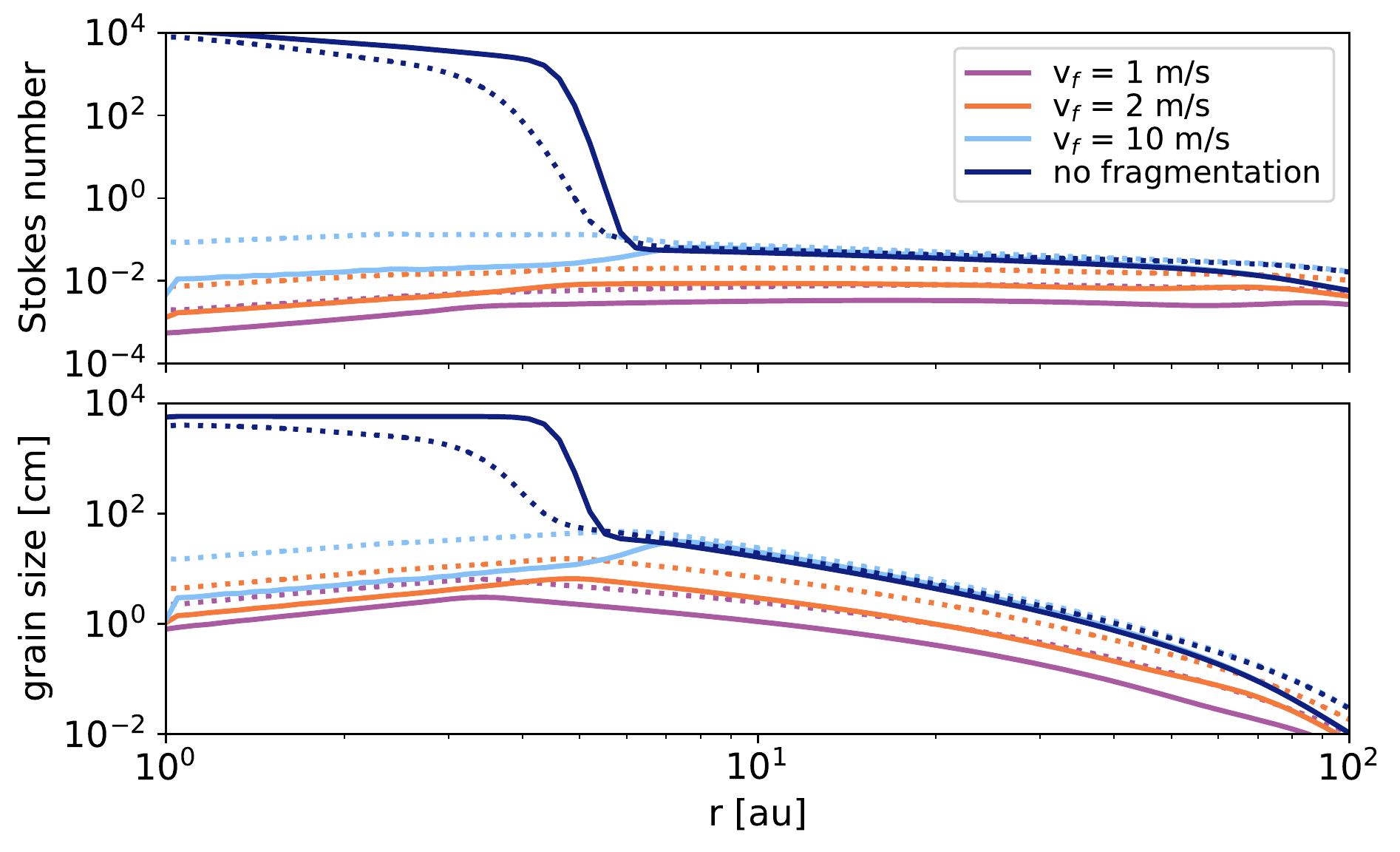}
\caption{Mass-averaged (solid lines) and flux-averaged (dotted lines) Stokes number ({\it{upper panel}}) and grain size ({\it lower panel}) after $2\cdot10^5$~years of evolution in the models of compact protoplanetary disk with $\alpha=10^{-4}$ and varying fragmentation threshold velocity $v_{\rm f}$.}
         \label{fig:frag_size}
\end{figure}

\begin{figure*}
    \includegraphics[width=\hsize]{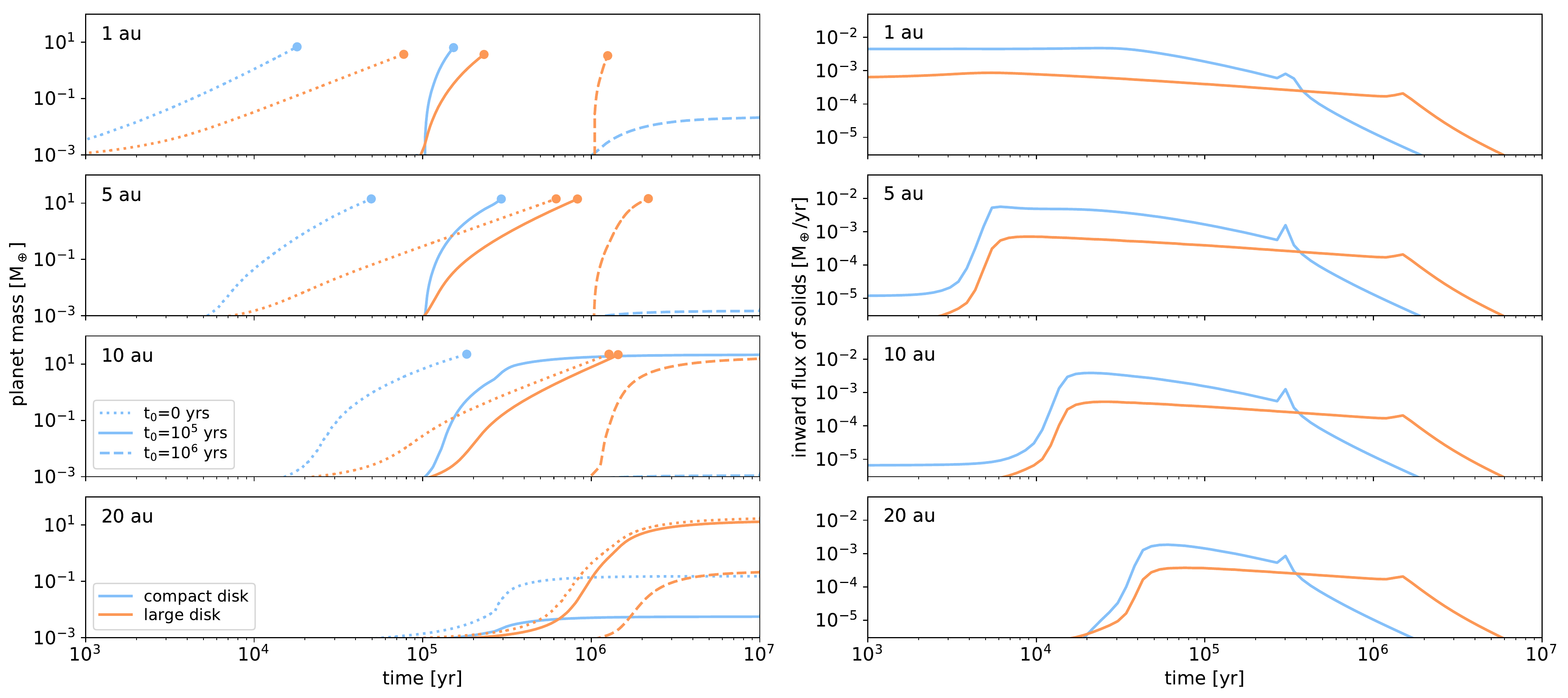}
     \caption{{\it Left:} Planetary embryos growth via pebble accretion in models with different characteristic radius, $\alpha=10^{-4}$, and the fragmentation threshold of v$_{\rm{f}}=10$~m/s. Lines ended with a circle represent growth tracks that reached the pebble isolation mass. {\it Right:} Inward flux of solids (integrated all dust sizes) at different locations in the same models.}
\label{fig:planets_disksize}
\end{figure*}

The runs with $v_{\rm{f}}=10$~m/s and no fragmentation produced very similar results except for the inner region of the disk ($r\leq5$~au). Interestingly, in this inner region of the disk, the growth of embryos introduced at $t_0=0$ and $t_0=10^5$~years proceeds faster in the runs with fragmentation than in the run without fragmentation. In the run without fragmentation, the embryo located at 1~au does not reach its pebble isolation mass irrespective on the introduction time, while in the run with $v_{\rm{f}}=10$~m/s, the innermost embryo reaches the isolation mass after only tens of thousands of years. The reason for the slow growth of the close-in embryos in the model without fragmentation is that pebbles break through the drift barrier and continue growing quickly over the Stokes number of unity, in the Stokes drag regime. This process has been previously described in the literature, see e.g. \citet{2008A&A...480..859B, 2010A&A...513A..79B, 2012ApJ...752..106O, 2014MNRAS.437.3037L, 2014IAUS..310..208D}.

The breakthrough only happens in the inner part of the disk, as the outer part is still dominated by the radial drift barrier. Figure~\ref{fig:frag_size} presents the average Stokes numbers and grain sizes as a function of radial distance at $t=2\cdot10^5$~years of evolution in the four models with different fragmentation threshold. It is worth noting that the flux-averaged dust size (and Stokes number) is typically a bit larger than the mass-averaged value, except for the grains that break through the $\rm{St}=1$ barrier. This is because the radial drift velocity increases up to $\rm{St}=1$ and then decreases again (see Eq.~\ref{eq:vr}). The grain size of $\sim$100~m reached in the inner region in the run without fragmentation is an effect of the upper limit of the size grid used in the dust coagulation calculation rather than a physical value. The large solids do not drift efficiently, which manifests as the low pebble flux at 1~au in the right panel of Fig.~\ref{fig:planets_fragmentation}. Since the grain growth is limited to the upper boundary of the size grid, those results should be treated as approximate, as the pebble flux would drop even more if the growth could proceed without a limit. 

Except for reducing the pebble flux, breaking through the $\rm{St}>1$ barrier has another negative impact on planet growth rate. The efficiency of pebble accretion significantly drops for such large dust aggregates (see Fig.~\ref{fig:epsPA}), as technically they do not count as pebbles anymore. Thus, even though the total mass of solids passing around the close-in embryo over the disk lifetime is essentially the same, in the run without fragmentation the planet does not reach the pebble isolation mass.

This result shows that some level of dust fragmentation may be beneficial for the planetary growth via pebble accretion. Thus, we performed additional runs, lowering the fragmentation speed even more, to $v_{\rm{f}}=2$~m/s and $v_{\rm{f}}=1$~m/s, the latter value being roughly consistent with the conclusions from laboratory work \citep{2010A&A...513A..56G}. As visible in the right panel of Fig.~\ref{fig:planets_fragmentation}, decreasing the fragmentation speed leads to pebble flux reduction. In the runs with the low fragmentation speed, grains stay at smaller sizes (see Fig.~\ref{fig:frag_size}) and drift slower, leading to lower but long-lasting mass flux, as depletion of small aggregates takes longer. This effect is particularly important if the planetary embryo takes a long time to form. In the left panel of Fig.~\ref{fig:planets_fragmentation}, one can see that if embryos are introduced at 1~Myr of dust evolution, they grow the most in the run with the lowest fragmentation threshold, $v_{\rm{f}}=1$~m/s. 

Interestingly, for embryos introduced at $t_0=10^5$~years, the $v_{\rm{f}}=2$~m/s is the most favorable value in terms of planetary growth. This is because, while the flux in the runs with the higher fragmentation threshold is already decreasing at that point, the flux in the $v_{\rm{f}}=2$~m/s run stays high for another couple of hundred thousands years (see the right panel of Fig.~\ref{fig:planets_fragmentation}). At the same time, the flux is factor of two higher in the $v_{\rm{f}}=2$~m/s run than in the $v_{\rm{f}}=1$~m/s case.
 
\subsection{Disk size}

In the results presented in Sect.~\ref{sub:frag}, the pebble flux lasts for at most 3~Myrs. However, millimeter and centimeter-sized pebbles are observed in protoplanetary disks older than several Myrs \citep{2005ApJ...626L.109W, 2010A&A...521A..66R}. \citet[][their Appendix~A]{2019A&A...622A.202J} noticed this problem and speculated that limited dust growth in the outer part of the disk may be the solution to retaining pebbles for long timescales. They have also shown that large initial disk size facilitates long-lasting pebble flux, a solution often used in pebble accretion models \citep{LJ2014, 2016A&A...589A..15S, 2018A&A...609C...2B}. While large disks are bright and thus relatively easy to image at high angular resolution \citep{2018ApJ...869L..41A}, there is evidence that most of the disks may, in fact, be small \citep{2018ApJ...859...21A, 2019ApJ...882...49L, 2019A&A...621A..76M, 2020ApJ...890..130T, 2020A&A...640A...5T}. In this section, we study the impact of disk size on the pebble flux and resulting planetary growth.

Figure~\ref{fig:planets_disksize} presents results of two models where the same initial mass was distributed either in a compact disk with the critical radius $r_c=30$~au or in a much larger disk with $r_c=300$~au. The other simulation parameters are the same: the fragmentation speed is $v_{\rm{f}}=10$~m/s and the turbulence strength is $\alpha=10^{-4}$. 

The compact disk produces a very high pebble flux, well over $10^{-3}$~M$_\oplus$/yr, which, however, only lasts for about $3\cdot10^5$~years and declines over the next couple of hundred thousands years (see the right panel of Fig.~\ref{fig:planets_disksize}). With such a high pebble flux, the embryos introduced at the beginning of the simulation take less than one hundred thousand years to reach their isolation mass. The embryo placed at 1~au reaches its isolation mass in little over ten thousand years, which is the fastest growth we observe in any of our runs. 

Pebble flux obtained in the large disk model is almost one order of magnitude lower than in the compact disk model, but lasts proportionally longer. Thus, even embryos introduced late during the simulation, at $t_0=10^6$~years, are able to obtain their isolation mass, at least inside of 10~au. Thus, extending the initial size of the disk gives similar results to lowering the fragmentation speed, as we discussed in the previous section. There is however, one important difference, which manifests itself in the planet growth results at and outside of 10~au. In the compact disk, a significant fraction of the mass is initially located inside of 10~au. The mass reservoir available for wide-orbit planets is significantly higher in the large disk than in the compact disk. Thus, for example the embryo introduced at $t_0=10^5$~years at 10~au, reaches its isolation mass in the large disk model but not in the compact disk model. 

With the large mass reservoir available even at wide orbital distances, it may be surprising that no embryo reaches its isolation mass outside of 10~au in the large disk model. This is because the pebble accretion efficiency for a constant embryo size significantly decreases with radial distance (see Fig.~\ref{fig:epsPA}). In this paper, we consider the initial embryo mass of 0.001~M$_\oplus$ irrespective of its location (and starting time). More massive embryos could probably still reach pebble isolation mass in the outer parts of the disk.

\subsection{Turbulence}\label{sub:alpha}

\begin{figure*}
\centering
\includegraphics[width=\hsize]{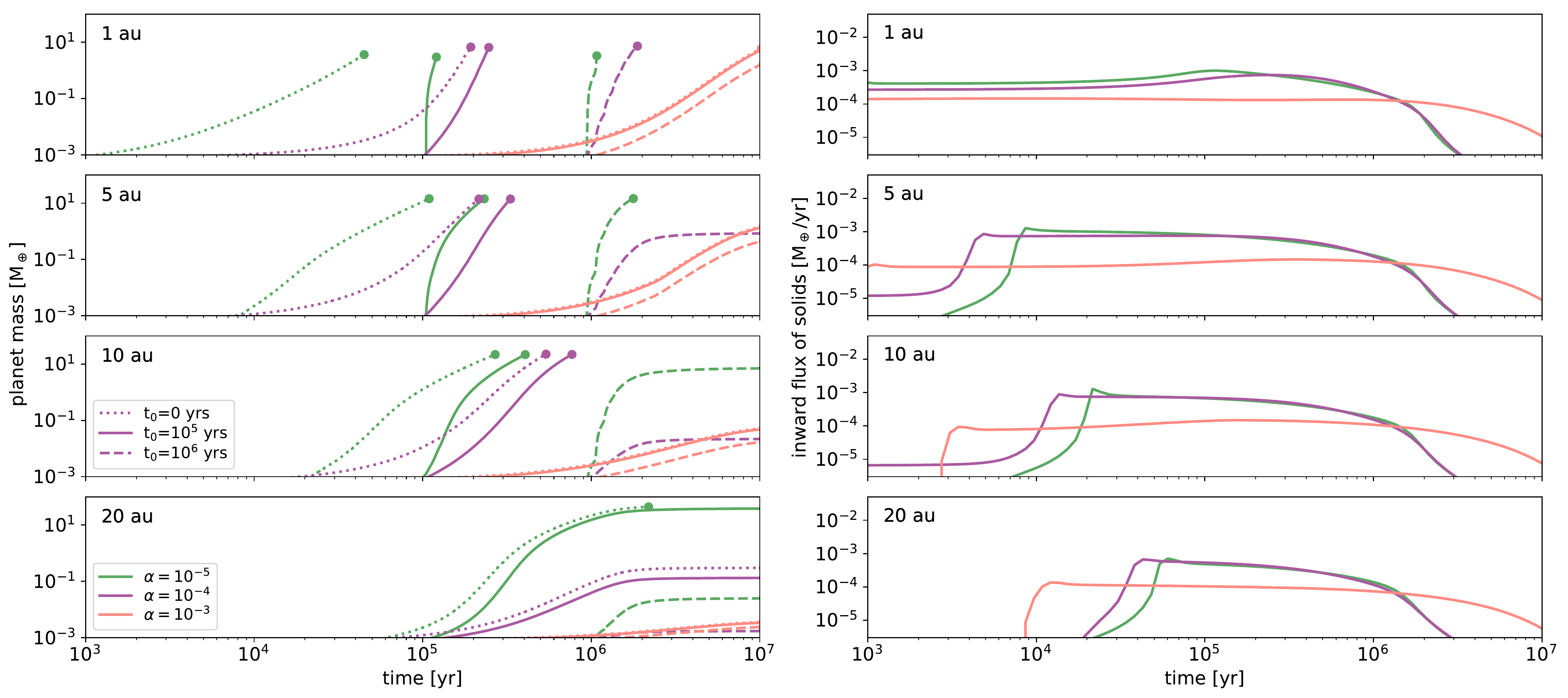}
\caption{{\it Left:} Planetary embryos growth via pebble accretion in the compact disk models with different turbulence strength parameter $\alpha$ and the fragmentation threshold of v$_{\rm{f}}=1$~m/s. Lines ended with a circle represent growth tracks that reached the pebble isolation mass. {\it Right:} Inward flux of solids (integrated all dust sizes) at different locations in the same models.}
         \label{fig:planets_alpha}
\end{figure*}

The turbulence level in protoplanetary disk is a key unknown. Observational constraints on the disk dispersal timescales suggest turbulence levels of $\alpha\approx10^{-2}$ if the dispersal would be driven solely by accretion driven by turbulent transport of angular momentum \citep{2007ApJ...662.1067H}. However, more recent disk models suggest that the turbulence level may vary within the disk and that large regions of the disk, particularly near to the midplane, may have low levels of turbulence \citep{2014A&A...566A..56L, 2016ApJ...821...80B}. This would be consistent with the disk images obtained in millimeter wavelengths, where the pebbles appear to be well-settled \citep{2016ApJ...816...25P, 2020A&A...642A.164V}, and also consistent with the low upper limits set by measurements of turbulent line broadening \citep{2015ApJ...813...99F, 2020ApJ...895..109F,2016A&A...592A..49T, 2018ApJ...864..133T}. Given the uncertainties and difficulties in constraining the turbulence level observationally, in this section, we consider the turbulence strength $\alpha$ as a free parameter.

Figure \ref{fig:planets_alpha} shows results of three runs performed on the backdrop of the compact disk model with different levels of turbulence. We keep the fragmentation speed at $v_f=1$~m/s in all the runs. As visible in the left panel of Fig.~\ref{fig:planets_alpha}, low turbulence promotes the planetary growth irrespective on the distance to the star and the starting time of the embryo. This has multiple reasons. First of all, lower turbulence level leads to better settling of the pebbles and higher pebble accretion efficiency \citep{2017ASSL..445..197O}. Lower turbulence speed promotes growth to larger pebble sizes, which increases the pebble flux, although this effect is stopped when the fragmentation starts to be dominated by the differential radial drift rather than turbulence (this happens for $\alpha \lesssim 10^{-4}$, see the right panels of Fig.~\ref{fig:planets_alpha}). 

It is worth noting that in case of the highest turbulence parameter value that we considered, $\alpha=10^{-3}$, no embryo reaches the pebble isolation mass. If the turbulence is strong, with $\alpha\geq10^{-3}$, the fragmentation threshold velocity needs to be much higher than $v_{\rm{f}}=1$~m/s, which we considered in models presented in Fig.~\ref{fig:planets_alpha}, to allow planetary embryos to benefit from pebble accretion. 

\section{Simple prediction of the pebble flux}\label{sub:predictor}

As we showed above, planetary growth by pebble accretion is very sensitive to disk parameters and details of dust evolution. The pebble flux depends on the initial mass budget, disk size, turbulence level, and fragmentation threshold speed. The pebble accretion efficiency is sensitive to pebble size and settling. The \texttt{DustPy} simulations described above are relatively expensive, typically taking several days of computations. In this paper, we propose a simple and efficient method that predicts the size and flux of pebbles for arbitrary protoplanetary disk without substructure, the \texttt{pebble predictor}\footnote{\texttt{pebble predictor} is publicly available at \href{url}{https://github.com/astrojoanna/pebble-predictor}. Version 1.0 of the script used in this paper is permanently available at \href{url}{https://doi.org/10.5281/zenodo.4383153}.}. 

The general mechanics of \texttt{pebble predictor} is closely related to \texttt{two-pop-py} presented by \citet{2012A&A...539A.148B}, albeit it a full time integration of the disk is not performed. The \texttt{pebble predictor} uses the initial state of the gas and dust disk surface density $\Sigma_{\rm{gas,0}}$ and $\Sigma_{\rm{dust,0}}$, temperature $T$, the turbulence strength $\alpha$, and the fragmentation threshold velocity $v_{\rm{f}}$ to predict the flux-averaged Stokes number and total flux of pebbles at every location within the disk and at any time point.

It is assumed that all the dust grains are initially in the form of micron-sized monomers and they grow in collisions driven by turbulence, which leads to the growth timescale of
\begin{equation}\label{eq:tgrowth}
    \tau_{\rm{growth}} = \frac{\Sigma_{\rm{gas,0}}}{\Sigma_{\rm{dust,0}}} \cdot \Omega_{\rm K}^{-1} \cdot \left(\frac{\alpha}{10^{-4}}\right)^{-1/3} \cdot \left(\frac{r}{\rm{au}}\right)^{1/3},
\end{equation}
where $\Omega_{\rm K}$ is the Keplerian frequency and $r$ is the radial distance to the central star.
The growth proceeds as an exponential function of time $t$
\begin{equation}
\mathrm{St_{ini}} = \mathrm{St}_0 \cdot \exp\left(\frac{t}{\tau_{\rm{growth}}}\right),
\end{equation}
where $\mathrm{St}_0$ is the Stokes number corresponding to the micron-sized monomers, until one of the growth barriers is encountered: fragmentation or radial drift. 
Fragmentation may be driven by turbulence or the differential radial drift. The maximum Stokes number of dust aggregates possible to achieve with respect to the turbulent fragmentation is 
\begin{equation}\label{eq:stfrag}
\mathrm{St_f} = \mathrm{f_f} \frac{v_{\rm f}^2}{3\alpha c_s^2},
\end{equation}
where $\mathrm{f_f}$ is a parameter, $v_{\rm{f}}$ is the fragmentation threshold velocity, $\alpha$ is turbulence strength parameter, and $c_s$ is the sound speed. If the radial drift speeds dominate over the turbulent speeds, the maximum Stokes number is 
\begin{equation}\label{eq:stdf}
\mathrm{St_{df}} = \mathrm{f_f} \frac{v_{\rm f}}{2v_\eta},
\end{equation}
where $v_\eta$ is the maximum drift speed (see~Eq.~\ref{eq:veta}). We adopt $\mathrm{f_f}=0.37$.

\begin{figure*}
\centering
\includegraphics[width=\hsize]{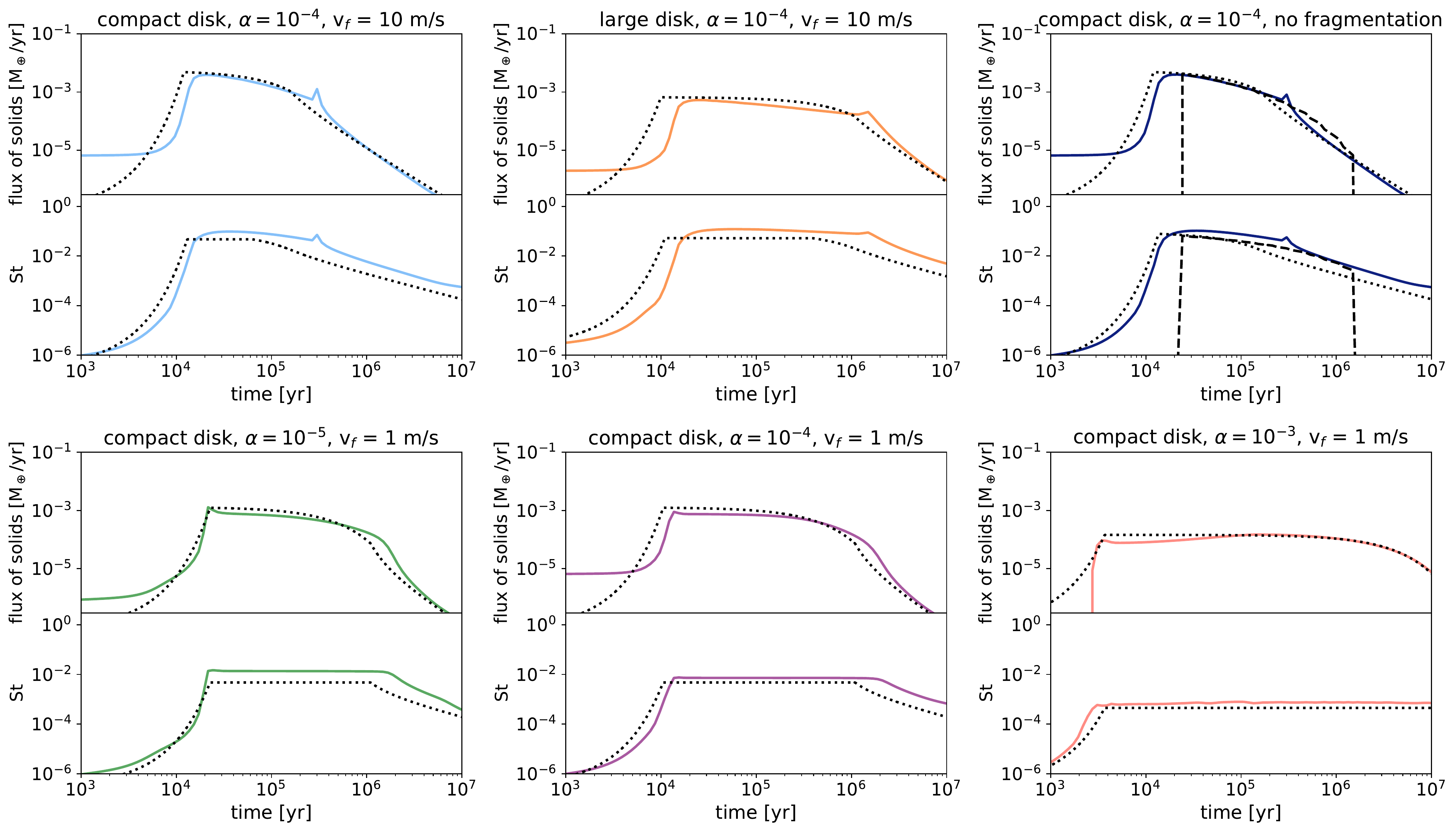}
\caption{Comparison of the pebble flux and Stokes number predictions and simulation results for different models, as indicated by the title of each subplot, at 10~au. The solid lines show the \texttt{DustPy} results, the dotted lines come from the \texttt{pebble predictor}. The dashed line in the top right panel corresponds to the model proposed by \citet{LJ2014}, which can only be used if the fragmentation is not included.}
         \label{fig:pebble_predictor}
\end{figure*}

We assume that the growth needs to be $\rm{f_{d/g}}=30$ times faster than drift in order for the grains to be not impacted by the radial drift \citep[see][their Sect. 4]{2012ApJ...752..106O}. We calculate the size of grains that are impacted by radial drift by checking the condition $\tau_{\rm{drift}} = \rm{f_{d/g}}\cdot\tau_{\rm{growth}}$, which leads to 
\begin{equation}\label{eq:std}
\mathrm{St_{drift}} = \frac{1}{\rm{f_{d/g}}\eta} \cdot \frac{\Sigma_{\rm{dust}}}{\Sigma_{\rm{gas}}},
\end{equation}
where $\eta = v_\eta/v_{\rm{K}}$ is the pressure gradient parameter.

The resulting, representative Stokes number $\it{St}$ in every radial bin is then chosen as a minimum of $\mathrm{St_{ini}}$, $\mathrm{St_f}$, $\mathrm{St_{df}}$, and $\mathrm{St_{drift}}$, however, it is not allowed to fall below its initial value $\mathrm{St_0}$. 

We use the first part of Eq.~(\ref{eq:vr}) to calculate the radial drift speed $v_r$ corresponding to the representative Stokes number and we can calculate the pebble flux as
\begin{equation}\label{eq:flux}
    \dot{m}_{p} = 2\pi r v_r \Sigma_{\rm{dust}},
\end{equation}
where $\Sigma_{\rm{dust}}$ would be the surface density of dust at a given time. This equation is however only correct if the grains grow faster than they drift. In such a case, there are always "enough" large grains, as the drift is depleting them at a slower rate than the growth could replenish them. In the opposite case, the pebble flux needs to be limited by taking into account that the supply of large grains can only be replenished at the growth timescale $\tau_{\rm{growth}}$. Thus we limit the inward drift velocity $v_r$ to 
\begin{equation}\label{eq:vrlimit}
    v_r = \min\left(\frac{2v_\eta \it{St}}{1+\it{St}^2}, \frac{r}{\rm{f_{d/g}} \tau_{\rm{growth}}}\right),
\end{equation}
which gives us a good estimate of the flux both in the fragmentation dominated and drift dominated regions of the disk. 

As mentioned above, \texttt{pebble predictor} does not perform a full time integration of the disk. We neglect the evolution of gas disk and we approximate the dust surface density evolution by keeping track of the mass budget. Using the initial $\Sigma_{\rm{dust,0}}$ profile supplied as an input, we calculate the initial mass $M_{\rm{out,0}}$ outside of every radial grid point. Using a time grid supplied as an input, \texttt{pebble predictor} calculates how much mass is left outside of every radial grid point at every time point $i$ using the values obtained in the previous time step:
\begin{equation}
 M_{\rm{out},i} = M_{\rm{out},i-1} - \dot{m}_{p,i-1} \cdot \left(t_i-t_{i-1}\right).
\end{equation}
The dust surface density used in Eqs~(\ref{eq:std}-\ref{eq:flux}) is then approximated as $\Sigma_{\rm{dust},i} = \Sigma_{\rm{dust,0}}\cdot M_{\rm{out},i}/ M_{\rm{out,0}}$. Since the time steps supplied in the input time grid are not necessarily fulfilling the CFL condition, this routine cannot be treated as a rigorous integration. Its goal is correcting the pebble flux for the remaining mass budget rather than recovering the actual surface density evolution. 

Figure~\ref{fig:pebble_predictor} shows the comparison of the \texttt{pebble predictor} and \texttt{DustPy} results for various setups varying the disk size, fragmentation speed, and turbulence level. The representative Stokes number from \texttt{pebble predictor} is benchmarked against the flux averaged Stokes number from the full coagulation calculation. On the top right panel of Fig.~\ref{fig:pebble_predictor}, we additionally show the flux and Stokes number prediction from the analytical model proposed by \citet{LJ2014}. In this model, it is assumed that the pebble flux is set by the initial dust surface density at the pebble formation front, which spreads radially with time. The general outcome of this model fits our results well, although the growth timescale is slightly underestimated and the pebble flux starts later than in the numerical results. The most significant difference reveals at later times, as the pebble flux in \citet{LJ2014} model abruptly stops when the pebble formation front reaches the outer edge of the disk. In reality, however, disk viscous spreading and dust diffusion modify the shape of the outer disk edge before the grains there reach pebble sizes. What is more, not all the pebbles at a given location grow and start drifting at once, which extending the duration of the pebble flux as well. Finding analytical prediction that would encompass all the relevant processes is not feasible. Thus, the \texttt{pebble predictor} is a semianalytic model relying on some empirical scaling.

The main limitations of \texttt{pebble predictor}, causing some discrepancies between its predictions and the full simulation results can be summarized as follows.
\begin{itemize}
    \item \texttt{pebble predictor} assumes that the growth is driven by turbulence (only one of the turbulent regimes calculated by \citealt{2007A&A...466..413O}, to be precise). This assumption breaks in the case of low turbulence, when the collisions are mostly driven by differential drift. For very small particles, the turbulent speeds fall into a different regime (see also the discussion in \citealt[][their appendix B]{2019ApJ...878..116P}), which leads to the need to modifying the growth timescale prescription with the empirical dependencies on turbulent strength $\alpha$ and radial distance in Eq.~(\ref{eq:tgrowth}).
    \item Only the Epstein drag regime is considered, which leads to error in estimating both the size and pebble flux where the growth enters the Stokes regime, so in the inner parts of disks in runs where dust fragmentation is not considered (see Sec.~\ref{sub:frag}).
    \item{Gas disk evolution is not included, leading to inaccuracies particularly at the outer regions of the disk, where disk spreading slows down the flow of pebbles.}
    \item{Similarly, dust advection with the gas flow is not considered in \texttt{pebble predictor}, which may lead to issues with estimating both the flux and pebble size correctly in runs when a significant fraction of solids is coupled to the gas, for example when the turbulence level is high and the fragmentation threshold is low.}
    \item{Diffusion is not considered, which is particularly important at the outer edge of the disk, where the dust-to-gas ratio gradient is strong \citep{2014ApJ...780..153B}. This leads to the characteristic peak in the dust flux visible in the runs involving compact disk and high fragmentation threshold, a feature which is not reproduced by the \texttt{pebble predictor}.}
    \item{Because the dust advection with gas flow and diffusion are not included, \texttt{pebble predictor} should not be applied to disks with substructure (see Sect.~\ref{sub:substructures}). A full model, such as \texttt{DustPy} or a more advanced simplified method such as \texttt{two-pop-py} developed by \citet{2012A&A...539A.148B} may be better choice in such case.}
\end{itemize}
Considering all the limitations, the \texttt{pebble predictor} does surprisingly well predicting the dependencies of pebble flux and size on the fragmentation speed, disk size, and turbulence level, as demonstrated in Fig.~\ref{fig:pebble_predictor}.

\section{Discussion}\label{sub:Discussion}

\begin{figure*}
\centering
\includegraphics[width=\hsize]{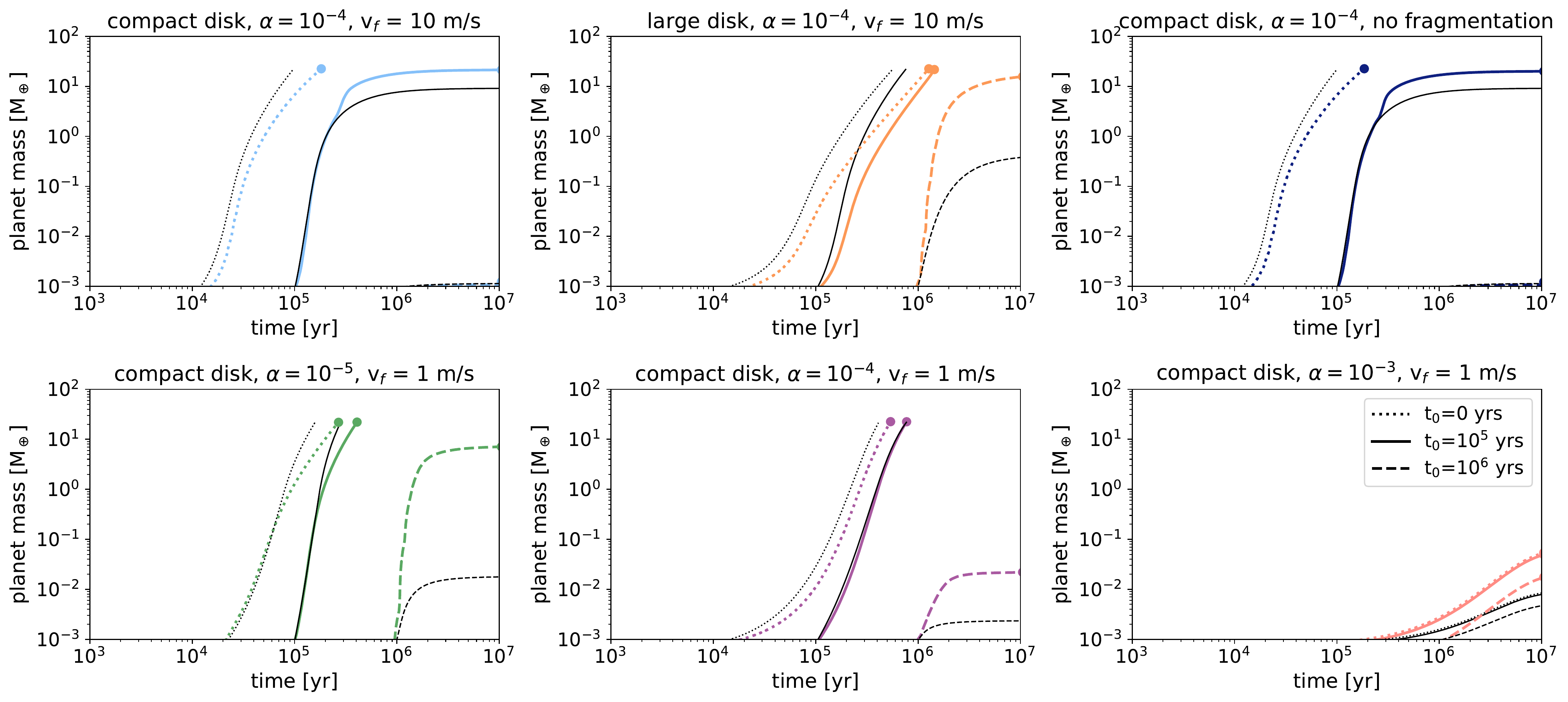}
\caption{Planetary embryos growth tracks obtained in various \texttt{DustPy} models (colored lines), as indicated by the title of each sub-plot, and calculated using the \texttt{pebble predictor} results (black lines). Planet growth is considered at 10~au with different starting times, as indicated by different line types.}
         \label{fig:planet_comp}
\end{figure*}

\subsection{Limitations of the (full) model}
In this paper, we coupled dust evolution calculated with the \texttt{DustPy} code and embryo growth by pebble accretion. Our models have certain limitations, some of them may impact the planetary growth results discussed in this paper. These are listed below.
\begin{itemize}
\item We do not consider that dust grains are composed of various materials and that evaporation of volatile components could cause decrease of the pebble flux in the inner regions of the disk.
\item We assume that all the dust aggregates are compact, with constant internal density, and neglect the effects of porosity. 
\item The structure of gas disk is independent of the evolution of solids. This is valid as long as the solids-to-gas ratio stays low, which is true in almost all of the models except for the one without fragmentation, when the grains break through the $\rm{St}=1$ barrier and pileup in the inner part of the disk. 
\item Our models correspond to one-planet-per-disk approach as we do not decrease the pebble reservoir by the amount accreted by each planet. Similarly, we do not include perturbations to disk structure caused by the growing planets. Such multiplanet effects could be important as many embryos would grow from the same pebble reservoir and the outermost embryo would block the pebble flux when it reaches its pebble isolation mass.
\item In this paper, we assume that the disk is fully turbulent, with a constant $\alpha$ value. Modern disk models predict that large parts of the disk may be free from turbulence and the accretion is primarily driven by magnetized disk winds \citep{2014A&A...566A..56L, 2017ApJ...845...75B}. In such disks, the viscous heating would not be relevant, leading to lower temperatures in the inner part of the disk than we assume in our models \citep{2019ApJ...872...98M}. Lower temperatures typically promote pebble accretion as the grains grow to larger sizes and are more settled. On the other hand, the pebble isolation mass is lower for colder disks.
\item Similarly, we do not include the gas disk dispersal by effects other than viscous accretion. Magnetic and photoevaporative winds could completely change the disk structure, particularly in the later phases of its evolution, which would impact the pebble fluxes and probably abruptly stop planetary growth by pebble accretion.
\end{itemize}

\subsection{Planet migration}

Perhaps the most significant limitation of our models is that we do not include planetary migration. Interaction between a sufficiently massive planetary core and the gas disk leads to radial migration \citep{1997Icar..126..261W}. A typical timescale of the inward migration for Earth-mass planet at 1~au is 10$^5$~years \citep{2015A&A...578A..36O}, comparable to the timescale of planetary growth by pebble accretion. 

Many authors have studied the interplay between the planetary migration and planet growth in pebble accretion scenario. \citet{2015A&A...582A.112B} showed that planets mostly migrate in the type I regime while accreting pebbles, but the migration direction may be both inward and outward, depending on their radial location. Planets that reach masses allowing for gas accretion typically outgrow the outward migration region and significantly change their radial location, by tens of au. Nevertheless, planet growth by pebble accretion is generally fast enough for planets to reach the pebble isolation mass before falling onto the central star due to migration \citep{2016A&A...590A.101B, 2018MNRAS.474..886N, 2019A&A...622A.202J}.

A connection between planet migration and pebble accretion including the self-consistent dust evolution model will be a subject to our future work.

\subsection{Validity of the single dust size approach in pebble accretion models}

In the models presented in this paper, we used dust evolution model that includes distribution of dust sizes. For every size bin, we use the solid flux calculated by the \texttt{DustPy} code, and multiply it by the pebble efficiency factor $\varepsilon_{\rm{PA}}$, corresponding to this pebble size, to get the pebble accretion rate. We sum up contribution from each size bin to calculate planet accretion rate, see Eq.~(\ref{eq:planetgrowth})\footnote{The sum in Eq.~(\ref{eq:planetgrowth}) may actually be omitted by using a single pebble accretion efficiency parameter calculated for the flux-averaged Stokes number and the total flux of solids. This approach gives a very good agreement with the results presented in this paper.}. However, since this type of model is computationally expensive, many authors employ simplified models of dust evolution. For example, the two population model for dust evolution proposed by \citet{2012A&A...539A.148B} is recently commonly used to compute dust evolution in protoplanetary disk as it allows us recovering the dust density evolution at a relatively low computational cost, without solving the full coagulation equation \citep[see. e.g.,][]{2016A&A...594A.105D, 2017MNRAS.465.3865C, 2018ApJ...863...97T, 2019A&A...627A..50C, 2020A&A...635A.149G}.

In this paper, we propose even simpler way of predicting pebble size and flux based on one representative grain size at every distance, the \texttt{pebble predictor} (see Sect.~\ref{sub:predictor}). It is justified to ask whether these simple models can be reliably used to calculate planetary growth via pebble accretion even though they do not calculate the full size distribution. We tested this by using the results of \texttt{pebble predictor} to calculate embryo growth and compared the results to the results obtained in the \texttt{DustPy} models in Fig.~\ref{fig:planet_comp}. Although there are some differences, the results in terms of the time of reaching the pebble isolation mass or the final embryo mass usually agree within one order of magnitude. The biggest differences are for planets introduced late, at $t_0=10^6$~years. This is because the \texttt{pebble predictor} does not include the gas disk evolution. 

\subsection{Early planet formation}

Studies of mass reservoirs of planet forming disks infer that planet(-esimal) formation should start early \citep{2011MNRAS.412L..88G, 2014MNRAS.445.3315N, 2018A&A...618L...3M, 2020arXiv200602812T}. There is growing evidence that planet formation process is indeed well underway in young disks \citep{2018NatAs...2..646H, 2020Natur.586..228S}. Pebble accretion may certainly be an avenue to fast planet formation, as young, massive disks should facilitate very high pebble fluxes.

Our results show that the pebble flux is the highest just after the dust aggregates at a given orbital distance reach their maximum size. Thus, if a large enough embryo forms early enough, it may benefits from this high pebble flux and reach its pebble isolation mass already during early stages of disk evolution \citep{2019MNRAS.484.1574T}. The pathway to such an early formation of massive planetesimals is not yet understood. The fastest route to planetesimal formation, the streaming instability, needs an enhanced solids-to-gas ratio to operate \citep{2009ApJ...704L..75J, 2010ApJ...722.1437B}. Enhancing their density requires significant redistribution of solids, which may take a long time \citep{2016A&A...594A.105D}. The cold finger effect at water snow line could lead to formation of some planetesimals already during the disk buildup stage \citep{2018A&A...614A..62D}. However, the planetesimals formed via the streaming instability are most likely not large enough to accrete pebbles and intermediate stage of planetesimal accretion is necessary \citep{2019A&A...624A.114L}.

\subsection{Substructures}\label{sub:substructures}

In this paper, we focused on the smooth disk model. However, there is observational evidence that substructures are ubiquitous in protoplanetary disks \citep{2015ApJ...808L...3A, 2018ApJ...869...17L, 2018ApJ...869L..41A}. Substructures are thought to modify the evolution of dust, allowing for solids retention over long timescales, consistent with observations \citep{2012A&A...538A.114P, 2019ApJ...878...39L}. They may be a preferential location of planetesimal and planet formation \citep{2003ApJ...583..996H, 2008A&A...487L...1B}. In particular, planetesimal formation may be triggered at the outer edge of a gap carved by a preexisting planet \citep{2019ApJ...884L...5S, 2020A&A...635A.110E, 2020A&A...644A..81S}.

In terms of pebble flux, substructures cause pileup of pebbles, likely speeding up local growth of planet located inside of the pressure bump  \citep{2020A&A...642A.140G, 2020A&A...638A...1M}. Existence of a long lasting substructure may be a way to enable growth of giant planets at large orbital distance. At the same time, however, overall pebble flux through the disk inward from a pressure bump would decrease as some, potentially significant, fraction of solids is held in the substructure.

Comprehensive knowledge of the substructure and detailed dust models are necessary to calculate the resulting pebble flux. One dimensional models of dust evolution in a pressure bump showed that dust growth can proceed to larger sizes inside of a pressure bump \citep{2008A&A...487L...1B, 2012A&A...545A..81P}. Dust trapping is hindered by diffusion and advection of small grains with the gas flow. There is a critical size of dust, which depends on the details of the bump structure, below which the grains cannot be trapped \citep{2016A&A...585A..35P, 2018ApJ...854..153W}. This effects can, in principle, be modeled with \texttt{DustPy}, which includes both dust diffusion and advection with the gas flow, however the results must be treated with caution, as recent two dimensional models show that the planet induced spiral wakes may modify the growth pattern \citep{2019ApJ...885...91D}. The simple flux prediction we proposed in Sect.~\ref{sub:predictor} does not include diffusion and grains advection with the gas flow and thus is not suitable for disks with substructures. We leave development of tools that would be useful in such disks for future work. 

\section{Summary}\label{sub:Summary}

In this paper, we studied planetary growth by pebble accretion in a protoplanetary disk without substructure. We investigated how the most uncertain parameters, the fragmentation velocity, disk size, and turbulence strength, impact the growth rate of embryos placed at different radial locations and times during the dust evolution. We found that planetary growth strongly depends on dust evolution and the pebble sizes and fluxes should be calculated self-consistently with the disk structure. An important byproduct of this paper is the \texttt{pebble predictor}, publicly available script that predicts the Stokes number and flux of pebbles in any smooth protoplanetary disk model for a given fragmentation threshold velocity at a very low computational cost (see Sect.~\ref{sub:predictor}).

One of our conclusions is that the planet growth by pebble accretion may be extremely fast if a large embryo is formed in the disk while the pebble flux is still high. Compact disk may produce very high pebble flux, leading to the close-in embryos reaching their pebble isolation mass within $10^4$--$10^5$~years, depending on the fragmentation speed, when the turbulence is not too strong ($\alpha\leq10^{-4}$). High turbulence levels universally reduce the planetary growth rate, while the effects of fragmentation threshold and disk size are not as obvious. 

Dust fragmentation keeps dust aggregates at pebble sizes, for which pebble accretion is the most efficient. Without fragmentation, dust growth would proceed over the pebble size limit, making the solids useless for growing planets by pebble accretion. Strong fragmentation keeps the pebbles small and decreases their drift speed, lowering pebble flux and slowing down the growth of embryos. At the same time, the low fragmentation threshold promotes keeping significant fraction of solids in the protoplanetary disk over long time and enables growth of embryos that formed late.

Compact disks are favorable for very rapid growth of close-in planets. In contrast, large disks are needed for formation of giant planets on wide orbits. Even with the large disk model, growing cores of giant planets outside of 10~au is challenging as the efficiency of pebble accretion drops with radial distance. One possible solution to this problem may be existence of dust traps that could facilitate planet growth by pebble accretion.\\

The authors acknowledge funding from the European Research Council (ERC) under the European Unions Horizon 2020 research and innovation programme under grant agreement No.~714769, the Deutsche Forschungsgemeinschaft (DFG, German Research Foundation) under Germany's Excellence Strategy – EXC-2094 – 390783311, and the support from the DFG Research Unit ``Transition Disks'' (FOR 2634/1, ER 685/8-1).

\bibliographystyle{aa.bst}
\bibliography{pebflux.bib}

\end{document}